\documentclass[journal,draftcls,onecolumn,12pt,twoside]{IEEEtranTCOM}
\normalsize

\usepackage{cite}

\ifCLASSINFOpdf
  \usepackage[pdftex]{graphicx}
  \graphicspath{ {..\pic} }
  \DeclareGraphicsExtensions{.pdf,.jpeg,.png}
\else
\fi

\usepackage[cmex10]{amsmath}
\usepackage{amssymb}
\newcommand{\Conv}{%
  \mathop{\scalebox{2}{\raisebox{-0.2ex}{$\circledast$}}
  }
}

\usepackage{algorithm, algpseudocode}
\usepackage[tight,footnotesize]{subfigure}

\usepackage{caption}
\begin{document}
%
\title{Channel Impulse Analysis of Light Propagation for Point-to-point Nano Communications through Cortical Neurons}
%
%
%

\author{Stefanus~Wirdatmadja, Josep Miquel Jornet, Yevgeni Koucheryavy,
 and~Sasitharan~Balasubramaniam
\thanks{S. Wirdatmadja, S. Balasubramaniam, and Y. Koucheryavy are with the Department of Electronic and Communication Engineering, Tampere University, Tampere, Finland, e-mail: \{stefanus.wirdatmadja, sasi.bala, evgeny.kucheryavy\}@tuni.fi.}
\thanks{J. M. Jornet is with the Department of
Electrical Engineering, University at Buffalo, The State University of New
York, NY 14260, USA, e-mail: jmjornet@buffalo.edu}
}

%
%

\markboth{IEEE Transactions on Communications}%
{Submitted paper}
%



\maketitle

\begin{abstract}
Recent Brain-Machine Interfaces have shifted towards miniature devices that are constructed from nanoscale components. While these devices can be implanted into the brain, their functionalities can be limited, and will require communication and networking to enable cooperation. One form of communication for neuron stimulation is the use of light. A number of considerations needs to be taken into account for the propagation and this includes diffraction, scattering, absorption, as well as attenuation. These properties are not only affected by the medium, but also by the cell's geometric shape. These factor affects both the direction and amplitude of the light wave. This paper analyzes the propagation path loss and geometrical gain, channel impulse and frequency response for light propagation along the neural tissue. The total attenuation depends on the propagation medium loss and geometrical gain, and the channel response is highly dependent on the quantity of cells along the path. Additionally, the optical properties of the medium also impacts on the time delay at the receiver and the width the location of the detectors. Based on the numerical analysis, spherical cells attenuate approximately 20\% of the transmitted power, which is less than the fusiform and pyramidal cells (35\% and 65\%, respectively).
\end{abstract}

\begin{IEEEkeywords}
nano communications, optogenetics, light propagation modelling, neural systems.
\end{IEEEkeywords}

%
\IEEEpeerreviewmaketitle

\section{Introduction}
\label{sec:intro}
%
%
%
%
\IEEEPARstart{I}{mplantable} medical devices in recent years have witnessed an exponential growth due largely to numerous supporting fields, including nanotechnology, computer science, and electrical engineering. The advancements in such fields are leading to miniature devices, constructed from biocompatible materials and powered by means of energy-harvesting systems, which can be permanently implanted. In this context, the emerging field of nano-communications is aimed at enabling the exchange of information and coordination between nano-devices. Two approaches for nano-communications have been developed in parallel, namely, nano-electromagnetic (EM) communications \cite{akyildiz2010electromagnetic} and molecular communications \cite{akyildiz2008nanonetworks}. A number of works have looked at molecular communications for neural systems. For example, \cite{abbasi2018controlled} conducts and analyzes in-vivo information transfer on the nervous system of the earthworm, \cite{ramezani2018impacts} proposes the complete synaptic communication channel model, and \cite{veletic2016upper} investigates the upper bound for neural synaptic communication. In the case of EM nano communication, a major challenge is the selection of appropriate frequency for signaling. that is relative to the size of the antenna components. This is because the reduction in size of an antenna means that the operating frequency also increases, possibly up into the terahertz (THz) band \cite{akyildiz2010electromagnetic}, and THz waves are unable to penetrate through biological tissues due to the high energy photons interacting with living cells at the molecular levels through the process of absorption \cite{johari2017nanoscale}. The implementation of THz communication in biological environment in nanoscale level is elaborated in \cite{elayan2017photothermal}. This means that other modes of communication are required in order to enable devices to communicate and network in biological tissues. While acoustic signals \cite{santagati2017software} have been proposed to allow devices to communicate, the unit circuitry may be larger than the envisioned micron-scale size of devices that need to be placed deep in a tissue. This challenge is further exacerbated when we consider implanting the devices in the cortex of the brain. 

\begin{figure}[t!]
    \centering
    \includegraphics[width=3.4in]{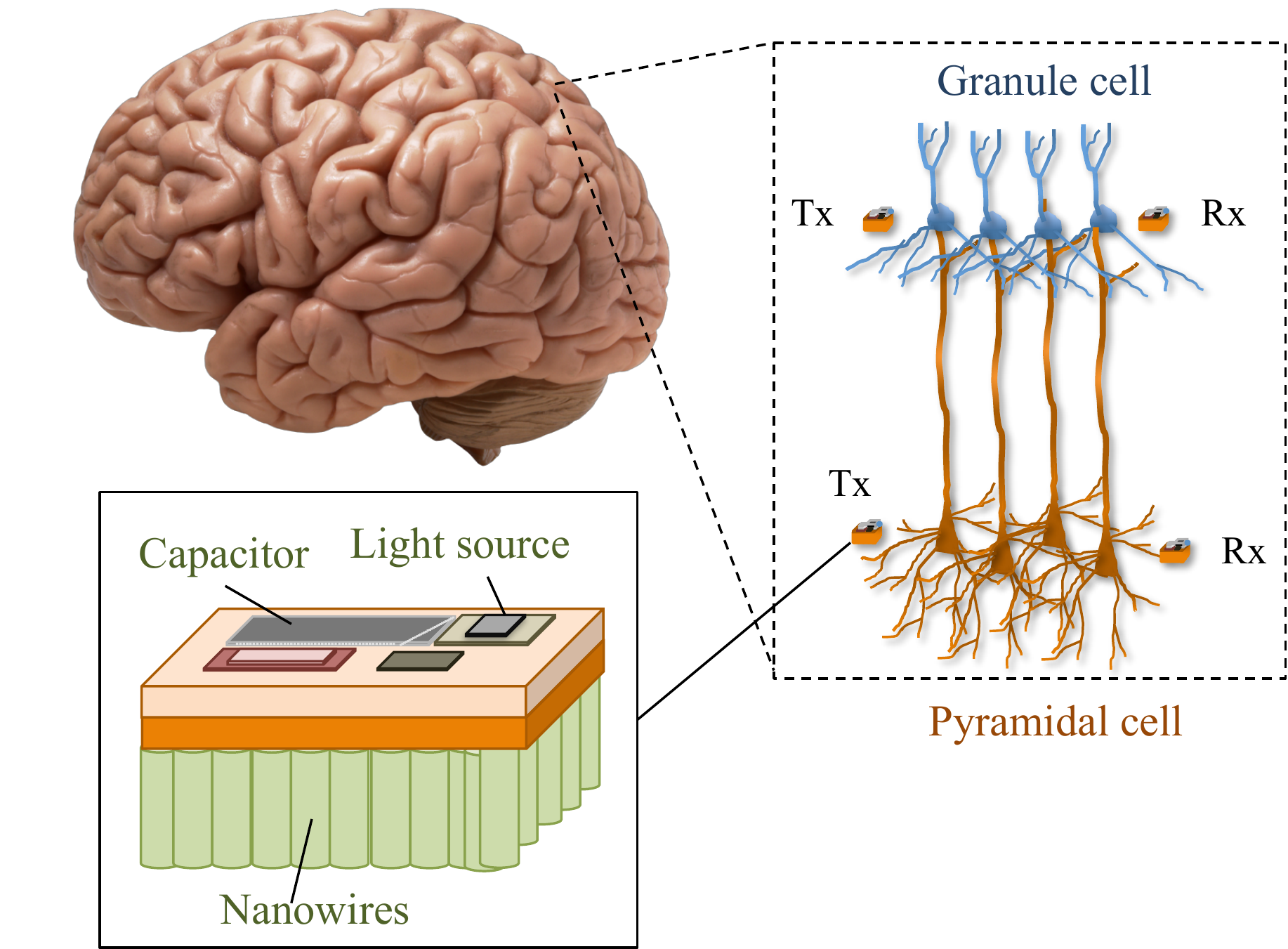}
    \caption{Illustration of the Wireless Optogenetic Nanodevice (WiOptND) network that are placed within the cortex. The communication between the devices is established using light that penetrates through the cells.}
    \label{fig:Overall}
\end{figure}

In \cite{wirdatmadja2017wireless} a miniature device, known as Wireless Optogenetic Nano Devices (WiOptND), for neural stimulation has been proposed. The device utilizes ultrasound signals for energy harvesting to produce power for a light source that is used for stimulating small population of neurons. Light has been used to interact with cells, and one example is the stimulation of neurons through the process called optogenetics. Some constraints in optogenetics stimulation are investigated in \cite{noel2018distortion} in terms of distortion as a spike generation delay. In other works, light has been used to communicate between devices through red blood cells \cite{johari2017nanoscale}. One key difference between THz waves and optical signals is that, while both are very high frequency EM waves, the latter do not suffer from absorption and are commonly found at the basis of nano-bio-sensing and imaging processes. While extensive modelling has been established on the properties and behaviour of the light for stimulating neurons, there have not been any solution towards how these devices can communicate with each other. 
In this paper, we propose communication between the WiOptND devices using light. As illustrated in Fig.~\ref{fig:Overall}, this could lead to miniature nanonetworks that are implanted into the brain cortex, and the communication and cooperation between the WiOptNDs can enable neural circuit stimulation of different micro-columns within the cortical cortex that have impaired connections. 

A major challenge with light communication through neural tissue, and in particular at micron-scale, is that the propagation of the light is largely determined by the physiological shape as well as the organelles within the cell. This is due to the power as well as size of the  source that produces light waves that are comparable to the size of the neuron. Therefore, the propagation behaviour of the light is largely dictated by the composition of the tissue. Neural tissue is composed of cells with different physiological properties, each of which contributes differently towards the propagation pattern. In this paper, we focus on the power delay profile, femto-pulse signal analysis in both time and Fourier domains, and the channel impulse response of light propagation for physical communication analysis in neural tissue. Our aim is to determine the channel impulse response of light signal as it propagates through different shapes as well as density of the neurons.  The contribution of this paper are as follows:

\begin{itemize}
  \item {\bf Geometric Analysis of Light Propagation and Path Loss Analysis:} We derive the total path loss formula for three different neuron geometry (fusiform, spherical, and pyramidal) based on optical properties of brain tissue and neuron. We analyze the focusing gain for multiple-(radial-based)-cell light propagation as the result of converging and diverging light phenomena through an array of cells.
  \item {\bf Numerical Analysis of Channel Impulse Response for Light Propagation:} We define a channel impulse response model based on the physiological shapes of three different neurons (e.g., fusiform, spherical, and pyramidal). This includes numerical analysis to determine the impact of light propagation through a line of neurons to determine how this impacts on the impulse response. The approach taken is through multiple ray tracings that simulates individual rays as they propagate through different number of cells. 
  \item {\bf Simulations of Ray Tracing Algorithm for Multiple-Cell-Propagation Path:} Simulations are conducted based on the pseudo code (Appendix A) for ray tracing through one-dimensional of cells. This simulation algorithm is based on recursion process for the light as it traverses through the last cell.
\end{itemize}

This paper is organized as follows: In Section \ref{sec:intro}, we present the background and motivation of this paper. In Section \ref{sec:pathloss}, we elaborate analytically the path loss characteristics of the light propagation in the brain. In both Section \ref{sec:cir} and \ref{sec:time_frequency}, we analyze the simulated system in both time and frequency domains. In Section \ref{sec:numerical_analysis}, we elaborate the results based on the simulated light propagation system and relate them with the geometrical and optical properties of the system. Finally, in Section \ref{sec:conclusion}, we give the conclusion based on the simulation and the overview of the optimum system.

\section{Path Loss Characteristics}
\label{sec:pathloss}
In general, for all cell geometries, the path loss is determined by the medium in which the light traverses. In our case, the medium consists of biological tissues and specifically neurons. Each propagation medium has three significant parameters that play a role in the behaviour of light transmission, namely absorption coefficient ($\mu_a$), reduced scattering coefficient ($\mu_s'$), and distance ($d$) between the transmitter and the receiver \cite{tuchin2015tissue} \cite{fang2007intrinsic}. The last parameter is heavily dependent on the refraction of light as it propagates through the medium, which is governed by the cell's geometry and the refractive index of different internal components ($n$) of the organelles.

\subsection{Beer-Lambert Law}

The propagation medium investigated in this paper consists of two biological mediums with different optical parameters. The two mediums are the brain interstitial fluids and each individual neuron. In this work, we consider that both remain constant throughout the propagation path. Therefore, the path loss analysis is based on the propagation distance, where the distance determines how the medium changes impact on the light path. The propagation path is analyzed using the modified \emph{Beer-Lambert law} and is represented as,
\begin{equation}
I(\lambda, d) = I_0(\lambda) e^{-\mu_a (\lambda) d DPF(\lambda, d)},
\end{equation}
where $I(\lambda)$ is light intensity at $\lambda$ wavelength on distance $d$, $I_o(\lambda)$ is the light intensity at the source, and $DPF(\lambda)$ is the Differential Path Length, which is the mean distance traveled by the light wave that is impacted by the shifted  direction due to the interaction with the neural tissue medium.  Based on this, the medium light transmittance $T(\lambda, d)$ can be written as
\begin{equation}
T(\lambda, d) = \frac{I(\lambda, d)}{I_0(\lambda)} =  e^{-\mu_a (\lambda) d DPF(\lambda,d )}.
\end{equation}

\subsection{Multiple Cell Path Loss}
\begin{figure*}[t!]
    \centering
    \includegraphics[width=\textwidth]{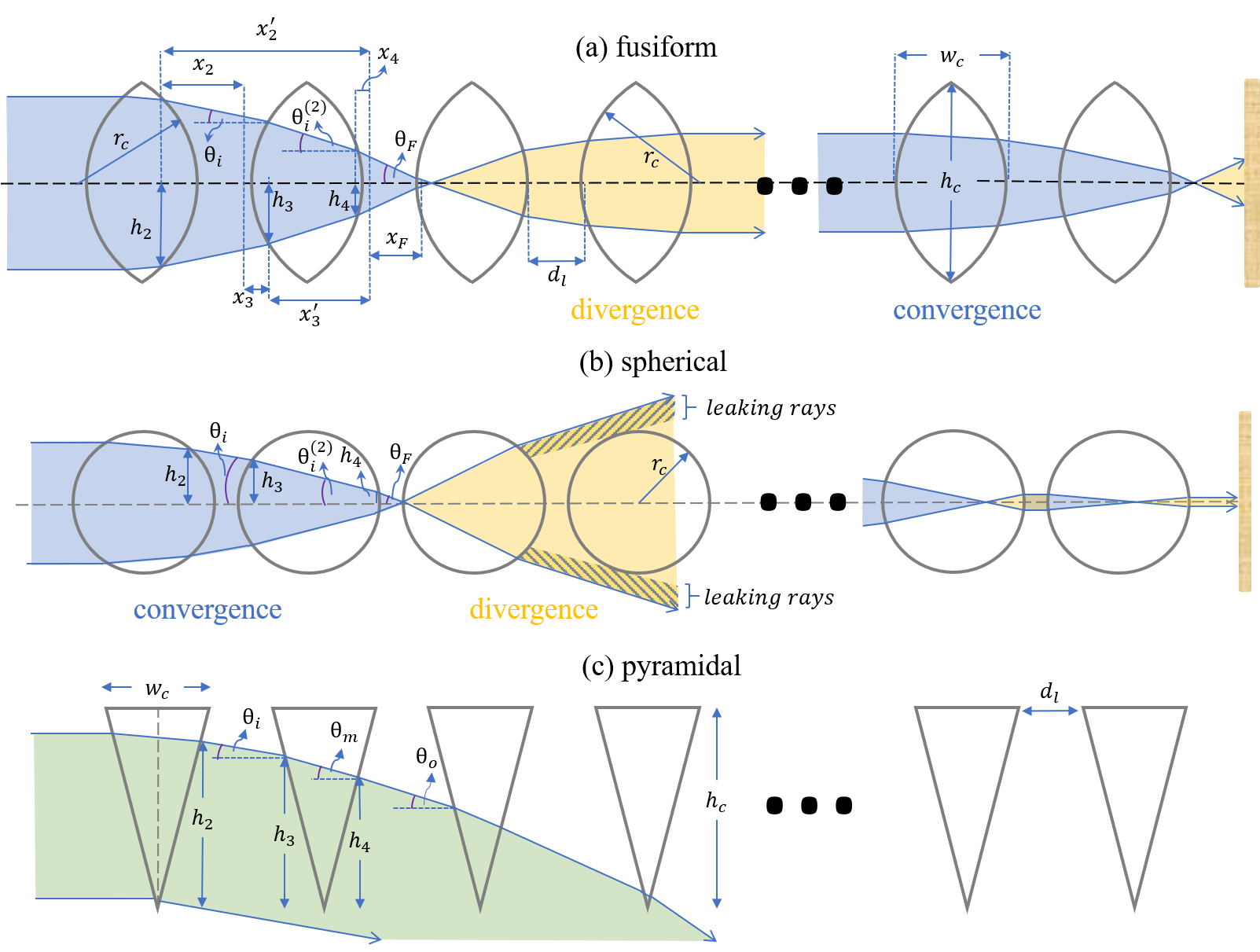}
    \caption{Geometrical analysis of light propagation using ray tracing as it propagates through a one-dimensional array of neurons, and this includes (a) fusiform, (b) spherical, (c) pyramidal cells.}
    \label{fig:geometry_propagation}
\end{figure*}

As discussed in Sec. \ref{sec:intro}, the geometry of the cell has a significant effect on the light propagation behaviour, and this is largely due to the size and aperture of the light source, and attenuation of the intensity. The geometrical analysis for the physiological shapes of the neurons is based on the models in \cite{wirdatmadja2019analysis}. In addition to the shape and size of each individual neuron, the light wave traverses a dense neural population. In this section, we elaborate the effect by the cell's geometry on the light propagation as it traverses a line of neurons that are of the same type. The general formula for the total path loss for $N$ number of any given shape of neurons is written as
\begin{align}
    PL_{total} = &4.343 \Big[ N \mu_a^{(c)} (\lambda) \overline{d_a} DPF(\lambda,\overline{d_a} ) \nonumber\\ 
    &+ (N-1) \mu_a^{(u)} (\lambda) \overline{d_e} DPF(\lambda,\overline{d_e} ) \nonumber\\
    &+ \mu_a^{(u)} (\lambda) (d_E+d_R) DPF(\lambda,(d_E+d_R) ) \Big],
\end{align}
where $\overline{d_a}$ and $\overline{d_e}$ are the average propagation distances in a cell and between two cells based on its shape, respectively, $d_E$ and $d_R$ are the distances of the light source from the first cell and the location of the receiver from the last cell, respectively. The superscript $(c)$ or $(u)$ indicates whether the parameter belongs to the cell or brain tissue. In the following subsection, we present the derivation for both $\overline{d_a}$ and $\overline{d_e}$ for three different neurons that is analyzed in this paper, which includes \emph{Fusiform}, \emph{Spherical}, and \emph{Pyramidal}. The shapes for each of the cells, as well as the geometric analysis of light propagation through the tissue is  illustrated in Fig.~\ref{fig:geometry_propagation}.

\subsubsection{Fusiform cell}
\label{subsubsec:fusiform}
A fusiform cell can be projected onto a 2D-plane as an oval or convex lens, as illustrated in Fig.~\ref{fig:geometry_propagation}(a). The dimension of the cell is represented with a height $h_c$,  width $w_c$, a surface curvature radius of $r_c$, and their relationship is represented as 
\begin{equation}
    r_c = \frac{{h_c}^2+{w_c}^2}{4{w_c}}.
\end{equation}
A single neuron of this shape has a focusing behaviour when collimated light propagates through the cell. To investigate the light propagation behavior for multiple neurons organized in a one-dimensional array, several fusiform cells are positioned in a sequence so that the collimated light propagates in a non-line-of-sight manner from the second cell. Applying this model to observe the focusing phenomena leads to another light behaviour which is the divergence effect. The divergence effect occurs when the focus point becomes shorter as the light propagates through consecutive fusiform cells. Due to the cell geometry, the focusing/converging and diverging phenomena occurs alternately, and this phenomena is illustrated in Fig.~\ref{fig:geometry_propagation}(a). Using geometrical analysis, the average propagation distances inside the fusiform cell $d_a$ and between the cells $d_e$ are formulated as
\begin{align}
    \overline{d_a} = &\frac{4}{h_c} \int_{\frac{r_c-h_c}{2}}^{\frac{r_c}{2}} \sqrt{{r_c}^2-x^2} - (r_c - \frac{w_c}{2})dx \nonumber \nonumber \\
    = &\frac{1}{6h_c} \Bigg[ 6h_c w_c-12{r_c}^2\arcsin\left(\frac{{r_c}-h}{2r_c}\right) \nonumber \\
    & \qquad + 3(h_c-{r_c})\sqrt{3{r_c}^2+2{h_c}{r_c}-{h_c}^2} \nonumber \\ 
    &\qquad \qquad+\left(2{\pi}+\sqrt{27}\right){r_c}^2-12{h_c}{r_c} \Bigg],
\end{align}
\begin{align}
    \overline{d_e} = &d_l + \frac{2}{h_c} \int_{\frac{r_c-h_c}{2}}^{\frac{r_c}{2}} \frac{w_c}{2} - \sqrt{{r_c}^2 - x^2} dx \nonumber \\ 
    = &d_l + \frac{1}{12h_c} \Bigg[ -6h_c w_c +12{r_c}^2\arcsin\left(\frac{{r_c}-h_c}{2r_c}\right) \nonumber \\
    & \qquad \qquad \; +\left(3{r_c}-3h_c\right)\sqrt{3{r_c}^2+2h_c{r_c}-{h_c}^2} \nonumber \\
    &\qquad \qquad \qquad +\left(-2{\pi}-\sqrt{27}\right){r_c}^2+24h_c{r_c} \Bigg].
\end{align}

\subsubsection{Spherical cell}
\label{subsubsec:spherical}
 Fig.~\ref{fig:geometry_propagation}(b) illustrates a spherical shaped neuron with radius $r_c$ projected onto a 2D space. The behaviour of the spherical shaped cell is similar to the fusiform cell, and this is from the focusing effects. However, depending on the density of the neuron, the distance between the cells $d_l$ has a significant role in \emph{leaking} light rays which does not occurs in multiple fusiform neural tissue. Leaking rays are the rays that do not propagate to the next adjacent cell. Therefore, they are not further transmitted along the propagation path. However, the alternating converging and diverging phenomena are similar for both spherical and fusiform cells. The light propagation for multiple spherical neurons in a 1-dimensional array is illustrated in Fig.~\ref{fig:geometry_propagation}(b). For the spherical cell the average propagation distance $d_a$ inside a cell and between cells $d_e$ is given by 
\begin{equation}
    \overline{d_a} = \frac{2}{r_c} \int_0^{r_c} \sqrt{r_c^2-x^2}dx = \frac{1}{2}\pi r_c,
\end{equation}
\begin{equation}
    \overline{d_e} = d_l + \frac{1}{r_c} \int_0^{r_c} r_c- \sqrt{{r_c}^2-x^2}dx = d_l -\frac{1}{4}(\pi-4){r_c}^2.
\end{equation}

Similar to the fusiform cell analysis, $\overline{d_a}$ and $\overline{d_e}$ of the spherical cell considers all the light rays along the propagation axis. However, the difference lies in the distance between the incoming and outgoing surfaces for the rays.

\subsubsection{Pyramidal cell}
\label{subsubsec:pyramidal}
The propagation behaviour for pyramidal cells is significantly different when compared to the two aforementioned neurons in terms of light ray traces, and this is illustrated in Fig.~\ref{fig:geometry_propagation}(c). A pyramidal cell tends to deviate the light path due to the refraction of two different medium, and takes on a behaviour that is very similar to a prism. In this case, the light traversing through multiple pyramidal cells deviates from its initial axis before it completely propagates from the  line path of the arrays of cells. The light deviation behaviour with respect to the cell's geometry is illustrated in Fig.~\ref{fig:geometry_propagation}(c). The average propagation distance $d_a$ inside a cell and between cells $d_e$ is represented as
\begin{equation} \label{eqn:pry_da} 
    \overline{d_a} = \frac{2}{h_c} \int_0^{h_c} \left[ \frac{w_c}{h_c}(x-h_c) + \frac{w_c}{2} \right] dx
    = \frac{1}{2} w_c,
\end{equation}
\begin{equation}\label{eqn:pry_de}
    \overline{d_e} = d_l + \frac{1}{h_c} \int_0^{h_c} \left[ \frac{w_c}{h_c}(x-h_c) + \frac{w_c}{2} \right] dx = d_l + \frac{1}{4}w_c.
\end{equation}
The 2-D projection of the pyramidal cell can be perceived as an isosceles triangle, which is similar to a prism. Therefore, the light ray deviation angle is governed by the medium refraction indices. Both \eqref{eqn:pry_da} and \eqref{eqn:pry_de} 
include ray traces from inside a cell and between two neighbouring cells.

\section{Channel Impulse Response}
\label{sec:cir}
The cell geometry and the tissue optical properties have a significant effect on the impulse response of the propagation channel. In this section, we derive how the geometrical analysis of the previous section plays a role on the  channel impulse response, and how this differs between the three different types of neurons. The combination of both the cell's geometry and the optical properties results in attenuation as well as delay of the light propagation to the detector. This is largely due to the collimation as well as divergence due to the geometric shape and results in multiple propagation paths of the light rays.

The general expression of multipath impulse response considering $N$ number of neurons that are placed in a 1-dimensional array can be formulated as \cite{rodriguez2013simulation}
\begin{equation}
\label{eq:total_cir}
    h(t; d_{total}, \lambda) = \Conv_{n=1}^{N} h_{a}^{(n)}(t; d, \lambda) \circledast \Conv_{n=1}^{N+1} h_{e}^{(n)}(t; d, \lambda),
\end{equation}
where $h^{(k)}(t; d, \lambda)$ represents the impulse response of the light ray corresponding to the $n^{th}$ cell and subscript \emph{a} or \emph{e} indicates if it is intracell or intercell propagation, $\lambda$ is the 
wavelength, and $t$ is the time. Furthermore, the intercell impulse response, $h_{e}^{(n)}$, consists of three elements, $h_{c}^{(n)}$, $h_{E}$, and $h_{R}$, and is represented as follows
\begin{equation}
    \Conv_{n=1}^{N+1} h_{e}^{(n)}(t; d, \lambda) = \Conv_{n=1}^{N-1} h_{c}^{(n)}(t; d^{(n)}, \lambda) \circledast  h_{E}(t; d_E, \lambda) \circledast h_{R}(t; d_R, \lambda)
\end{equation}
where subscripts \emph{c, E}, and \emph{R} indicates the propagation paths between the cells, light source, and the receiver (detector), respectively, $d_E$ is the distance between the light source and the first cell, and $d_R$ is the distance between the last cell and the receiver. As shown in \eqref{eq:total_cir}, the impulse response is based on the convolution as the light propagates through each individual cell along the path, and the brain tissue between the cells. 

The light source applied in this system is considered as a collimated light where the generated rays has equal intensity following the uniform distribution $\mathcal{U}(0, 1)$. In order to use the ray tracing model, the infinite rays should be discretized to K rays. Since the intensity parameter is used, the discretization process has no impact on the intensity value. To further elaborate on this, each intra and intercell propagation impulse response can be divided in two parts, namely, the attenuation component and the delay component. The attenuation follows the modified Beer-Lambert equation, while the delay can be expressed as a delta dirac function. Thus, the impulse response for $k^{th}$-path is given by
\begin{equation}
    h_{a}^{(k)}(t; d^{(k)}, \lambda) = I_E T_a^{(k)}(\lambda) \delta{(t-t_a^{(k)})}
    \circledast I_E T_e^{(k)}(\lambda) \delta{(t-t_e^{(k)})},
\end{equation}
where $I_E$ is the intensity emitted by the light source and $T^{(k)}$ is the transmittance of the $k^{th}$-path. Furthermore, the complete calculation of the impulse response for the K discretized ray paths traversing N-cells can be elaborated by a matrix represented as follows
\[
\begin{bmatrix}
    h_{a}^{(1)} \\
    h_{e}^{(1)} \\
    h_{a}^{(2)} \\
    h_{e}^{(2)} \\
    \vdots \\
    h_{a}^{(K)} \\
    h_{e}^{(K)} \\
\end{bmatrix}
= I_E \cdot e^{\underline{\eta}^T}
\begin{bmatrix}
    \delta(t-\frac{|| \underline{d_a^{(1)}} ||_1}{v_a}) \\
    \delta(t-\frac{|| \underline{d_e^{(1)}} ||_1}{v_e}) \\
    \delta(t-\frac{|| \underline{d_a^{(2)}} ||_1}{v_a}) \\
    \delta(t-\frac{|| \underline{d_e^{(2)}} ||_1}{v_e}) \\
    \vdots \\
    \delta(t-\frac{|| \underline{d_a^{(K)}} ||_1}{v_a}) \\
    \delta(t-\frac{|| \underline{d_e^{(K)}} ||_1}{v_e}) \\
\end{bmatrix},
\]
where $\mu_a$ is the absorption coefficient and the superscript $(c)$ or $(u)$ indicates the neuron or brain tissue, respectively, $\eta$ represents the matrix product of the distance and the DPF, $v$ is the light velocity in a medium, and subscript $a$ or $e$ indicates if the path is intra or intercell propagation, respectively. The velocity value can be obtained by $v=\frac{c}{n}$, where $c$ is the light speed in vacuum and $n$ is the refractive index of the medium. The matrix product $\eta$ is further derived as
\[
\underline{\eta} =
\begin{bmatrix}
    \eta_{a}^{(1)} \\
    \eta_{e}^{(1)} \\
    \eta_{a}^{(2)} \\
    \eta_{e}^{(2)} \\
    \vdots \\
    \eta_{a}^{(K)} \\
    \eta_{e}^{(K)} \\
\end{bmatrix} =
\begin{bmatrix}
    \sum_{n=1}^N d_a^{(1,n)} \\ 
    \sum_{n=1}^{N+1} d_e^{(1,n)} \\
    \sum_{n=1}^N d_a^{(2,n)} \\
    \sum_{n=1}^{N+1} d_e^{(2,n)} \\
    \dots \\
    \sum_{n=1}^N d_a^{(K,n)} \\
    \sum_{n=1}^{N+1} d_e^{(K,n)}
\end{bmatrix}^T
\begin{bmatrix}
    DPF_a^{(1)} \\
    DPF_e^{(1)} \\
    DPF_a^{(2)} \\
    DPF_e^{(2)} \\
    \vdots \\
    DPF_a^{(K)} \\
    DPF_e^{(K)}
\end{bmatrix}.
\]

The DPF is the element which is included in the modified \emph{Beer-Lambert} law and is affected by the optical medium properties, namely the absorption ($\mu_a$) and the reduced scattering ($\mu_s'$) coefficients \cite{scholkmann2013general}, and is represented as follows 
\begin{equation} \label{eq:dpf}
    DPF^{(k)}(\lambda) = \frac{1}{2}\bigg( \frac{3 \mu_s'(\lambda)}{\mu_a(\lambda)}  \bigg)^{1/2} \bigg[ 1- \frac{1}{1+d (3\mu_a(\lambda) \mu_s'(\lambda))^{1/2}} \bigg].
\end{equation}
The propagation channel elaborated in this paper consists of two medium (cell and brain tissue) which are categorized as intra and intercell propagation medium. Therefore, the DPF equation \eqref{eq:dpf} should be applied to all paths with respect to its medium.

\section{Frequency domain analysis}
\label{sec:time_frequency}
Our analysis is based on the same setup that is used in an optogenetic system. In optogenetics, the wavelength that is used for the neuron stimulation is based on the  visible 450-480-nm blue light, and this is the same wavelength that is used for the light communication between the WiOptND devices in our proposed model. The communication is established through light propagation that is represented a as short Gaussian shaped pulse. The Gaussian shaped pulse is the product of a cosine function and a Gaussian envelope function. For a light wave, the Gaussian pulse can be expressed as \cite{rulliere2005femtosecond}
\begin{equation}
\label{eq:E_t}
   E_t = Re\{E_0 e^{-4 ln(2)\big(\frac{t}{\tau}\big)^2 + i \omega_0 t}\}
\end{equation}
where $E_t$ and $E_0$ are the electric field with respect to time and at $t=0$, respectively, $\omega_0$ is the angular frequency of the light wave, and $\tau$ is the Full-Width at Half-Maximum (FWHM) pulse duration. On the receiver side, the time delay is added to the pulse waveform and it is formulated as
\begin{equation}
\label{eq:E_r}
   E_r = Re\{\gamma E_0 e^{-4 ln(2)\big(\frac{t-t_d}{\tau}\big)^2 + i \omega_0 (t-t_d)}\},
\end{equation}
where $t_d$ denotes the time delay caused by the propagation path and $\gamma = \big (\frac{r_E}{r_D} \big)^2$ is the area (proportional to the square of radius) ratio due to focusing effect, $r_E$ is the radius of the light source and $r_D$ is at the detector \cite{wirdatmadja2018light}.
The result from convolution series of channel impulse response presented in \eqref{eq:total_cir} can also be obtained by analysing the Fourier transforms of the transmitted signal, $\mathcal{F}(E_t)$, and the received signal $\mathcal{F}(E_r)$. The channel impulse response can be obtained by applying the inverse Fourier transform of the division, and is represented as
\begin{equation}
    h(t;d,\lambda) = \mathcal{F}^{-1} (H(f;d, \lambda) = \mathcal{F}^{-1}\Bigg( \frac{\mathcal{F}(E_r(t;d,\lambda)}{\mathcal{F}(E_t(t;d,\lambda)} \Bigg).
\end{equation}

Based on ~\eqref{eq:E_t} and \eqref{eq:E_r}, we can observe that the input and output relationship is heavily dependent on the delay caused by the propagation medium characteristics.

\section{Numerical Analysis}
\label{sec:numerical_analysis}
In this paper, the light propagation is simulated using MATLAB, where a function created that generates the ray geometrical propagation parameter inside each neuron and after the cell. Our approach used for the light propagation modeling is based on ray tracing. Algorithm~\ref{alg:ray_function} presents an example for the fusiform cell ray tracing process. The function for fusiform and spherical cells is similar due to their geometrical similarities. This function is iteratively executed and combined with the ray tracing by applying the focusing parameter $\gamma$ from \eqref{eq:E_r}, which is determined by the illuminated detection area. Table~\ref{table:init_var} lists all the parameters for the MATLAB simulation.

\begin{table}[ht]
\caption{Simulation Parameters}
\label{table:init_var}
\centering
\begin{tabular}{|c|c|l|}
 \hline
 Parameter & Value [Unit] & Description \\[1ex]
 \hline
  \hline
  $\lambda$ &   $456$ [$nm$]  & Visible blue light wavelength\\
   \hline
 $n_c$ & $1.36$ & Refractive index of the cell \cite{levinson1926refractometric}\\
  \hline
  $n_t$ & $1.35$ & Refractive index of the tissue\\
   \hline
 $\mu_a^{(c)}$ & $0.9$ [$/mm$]  & Cell absorption coefficient \cite{yaroslavsky2002optical}\\
  \hline
 $\mu_s'^{(c)}$ & $3.43$ [$/mm$]  & Cell reduced scattering coefficient\\
  \hline
 $\mu_a^{(u)}$ & $1.34$ [$/mm$]  & Tissue absorption coefficient \cite{bosschaart2014literature}\\
  \hline
 $\mu_s'^{(u)}$ & $3.43$ [$/mm$]  & Tissue reduced scattering coefficient\\
  \hline
 $\tau$& $1$ [$fs$] & FWHM pulse duration \\
  \hline
\end{tabular}
\end{table}

\subsection{Medium Loss and Geometrical Gain}
\label{subsec:loss_and_gain}
\begin{figure}
\centering
\begin{minipage}{.5\textwidth}
  \centering
  \includegraphics[width=.9\linewidth]{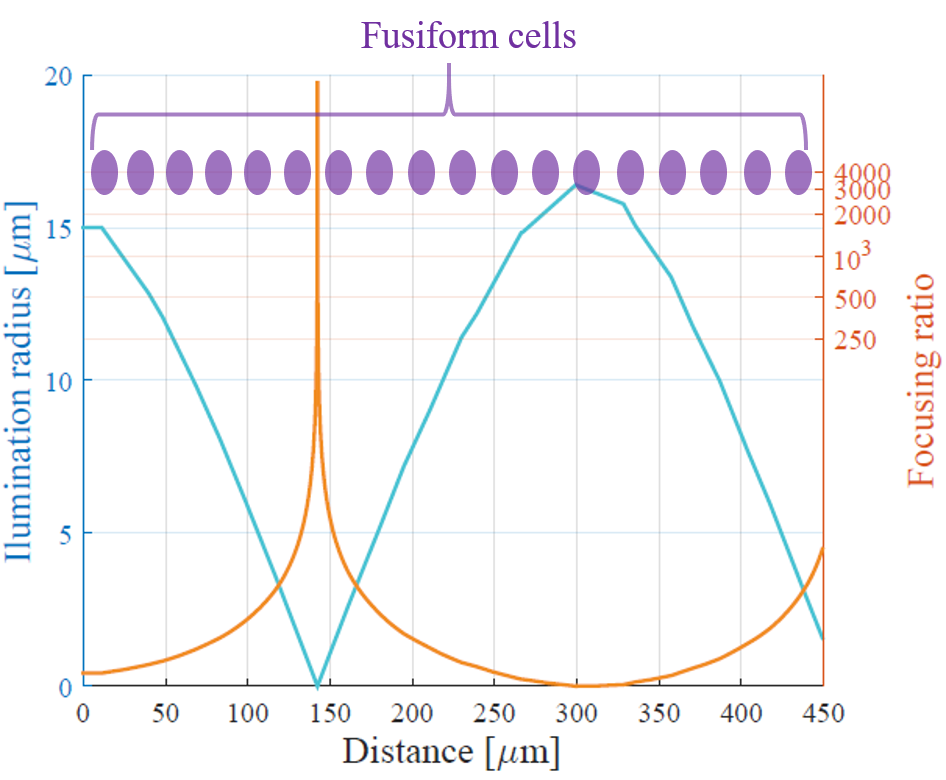}
  \captionof{figure}{Illumination radius (h$_{2}$/h$_{3}$/h$_{4}$) and\\ focusing ratio $\gamma$ for eighteen fusiform \\ neurons.}
  \label{fig:fusiform2}
\end{minipage}%
\begin{minipage}{.5\textwidth}
  \centering
  \includegraphics[width=.9\linewidth]{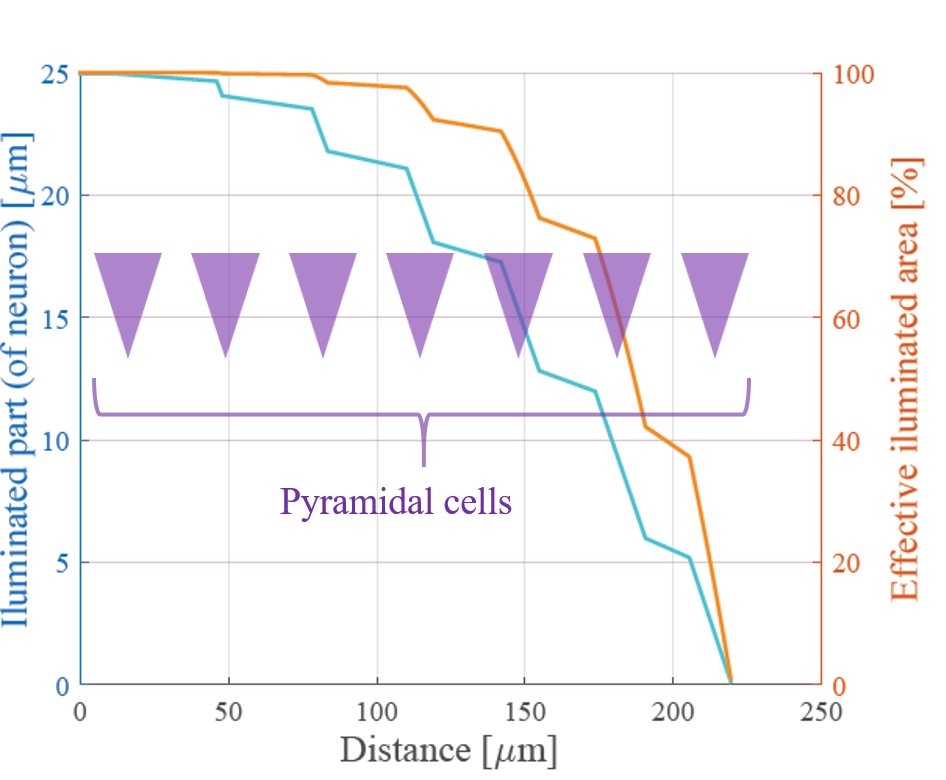}
  \captionof{figure}{Illumination height (h$_{2}$/h$_{3}$/h$_{4}$) and \\effective illumination area for seven \\ pyramidal neurons.}
  \label{fig:pyramidal2}
\end{minipage}
\end{figure}

The light wave traversing in the biological tissue experiences attenuation as discussed in Sec.~\ref{sec:pathloss}. The attenuation is mainly due to the optical properties of all the biological components in the cell medium. However, the fusiform and the spherical cells focuses the light rays as it enters into the cytoplasm, and this is due to the changes in the refractive index. This focusing effect is further increased when the light propagates into the nucleus, and once again this is due to the differences in the refractive indexes of the medium. Fig.~\ref{fig:fusiform2} shows the effect of the focusing ratio and illumination radius which contribute in the overall gain of the light intensity. This can be observed in the rise of the illumination as the light propagates through certain cells, and then divergence occurs leading to reduction in the illumination.

\begin{figure}[t!]
    \centering
    \includegraphics[width=0.8\textwidth]{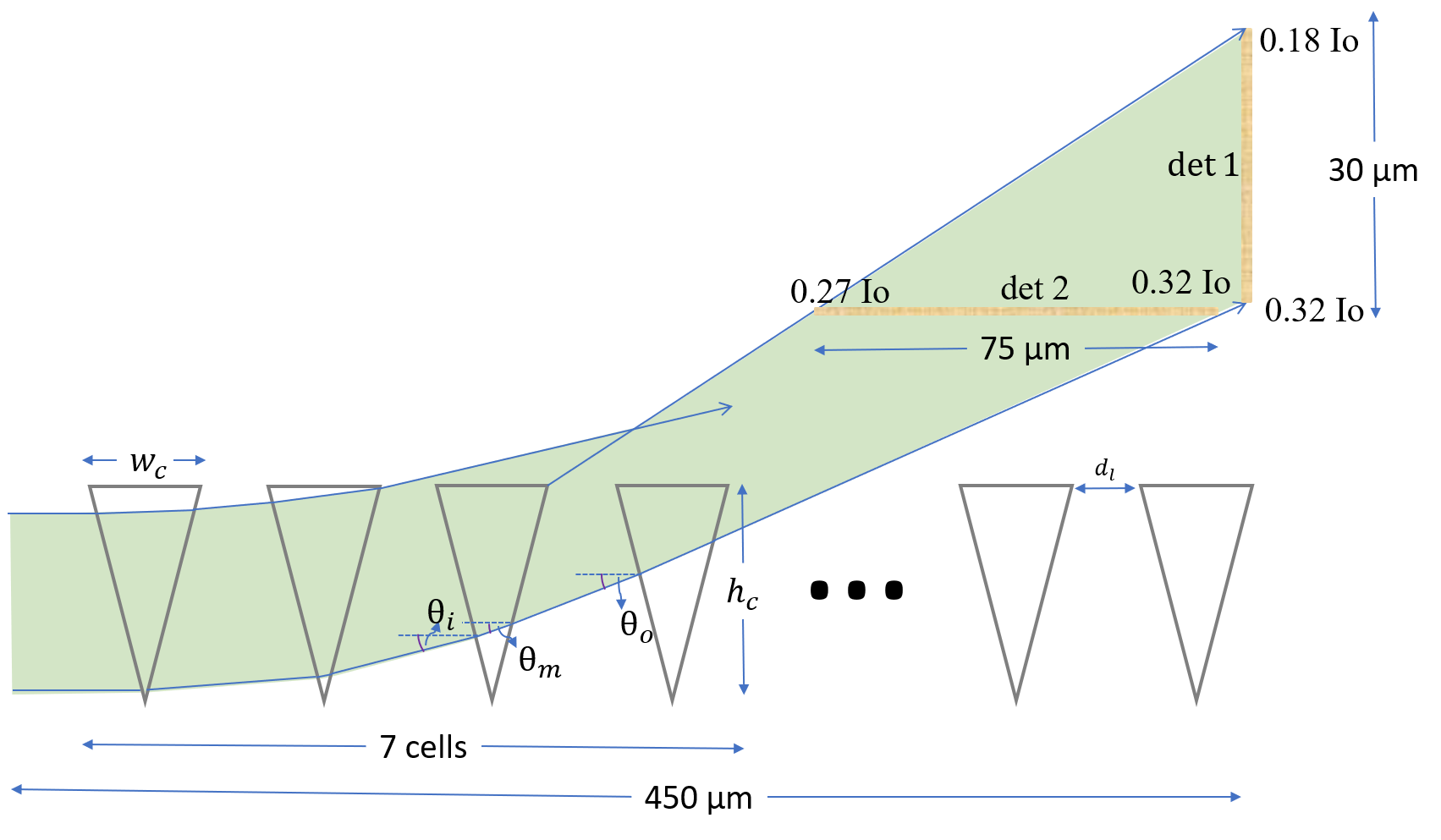}
    \caption{Light path deviation along an array of pyramidal cells due to the geometric shape.}
    \label{fig:ray_deviation}
\end{figure}

The focusing behaviour is not found in the pyramidal geometry. Fig.~\ref{fig:pyramidal2} presents the illumination and shows that it gradually reduces due to the divergence of the light path away from the line of cells as described above. This is solely due to the geometrical structure of the cell, as illustrated in Fig.~\ref{fig:ray_deviation}.

\begin{figure}[t!]
    \centering
    \includegraphics[width=3.2in]{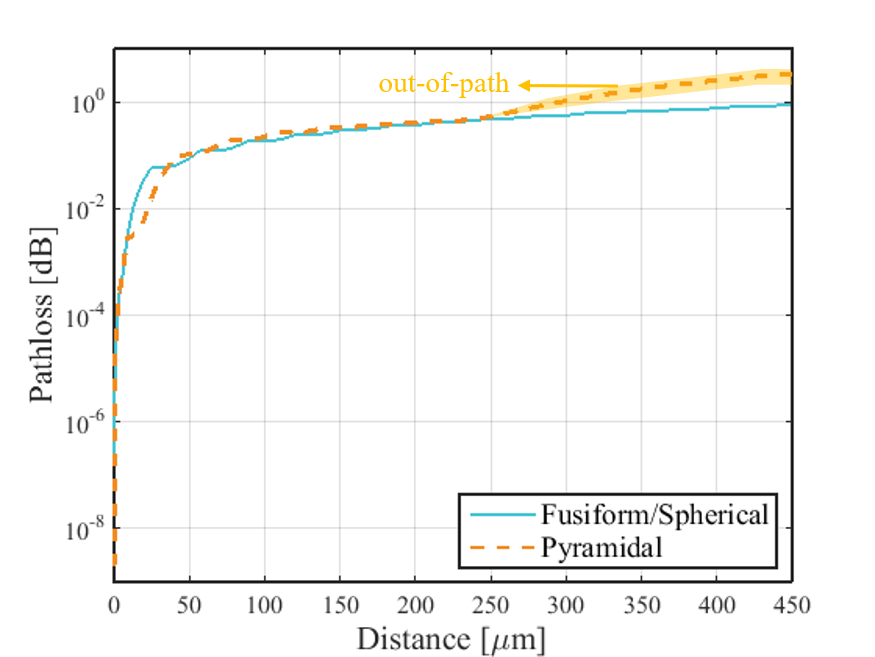}
    \caption{Light propagation path loss for the distance of 450 $\mu$m between the transmitter and receiver, with 18 cells in between.}
    \label{fig:pathloss_fus_pyr}
\end{figure}

Fig.~\ref{fig:pathloss_fus_pyr} shows the path loss within the brain tissue for the three different neurons, where the transmitter and receiver is separated by 450 $\mu$m. In case of the pyramidal cells, the ray deviation due to the geometrical refraction causes the gradient change (marked by yellow shade). The deviation indicates that the light does not penetrate through the remaining neurons along its path, and this is because for the fusiform cells, there are eighteen neurons along the propagation path, while the pyramidal cells has seven neurons due to the light path divergence.

\begin{figure*}[t!]
    \centering
    \includegraphics[width=7in]{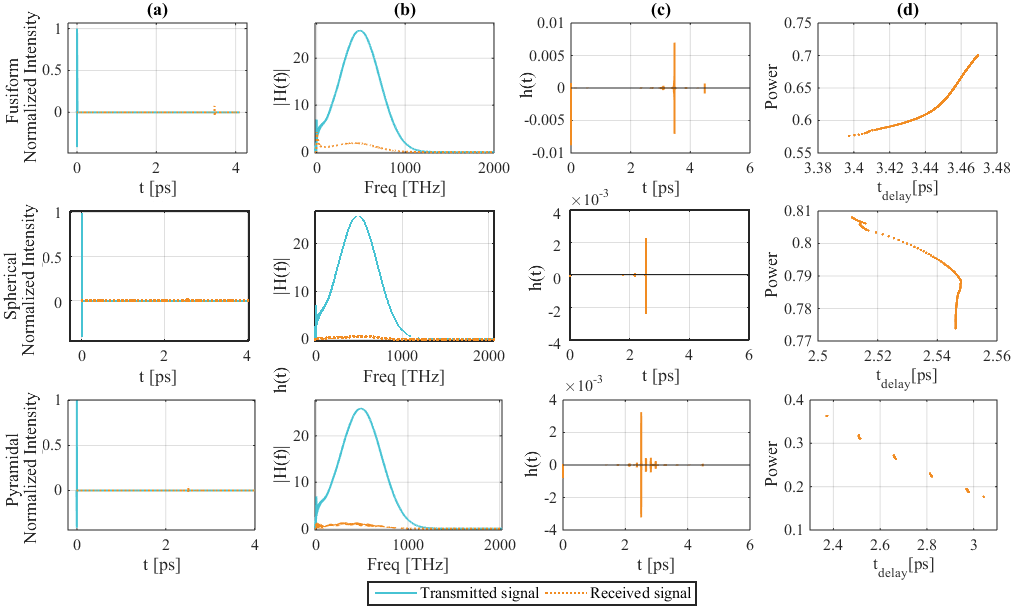}
    \caption{Time and frequency analysis of light propagation along fusiform, spherical, and pyramidal cells. (a) The normalized intensity of the transmitted and the received signals in the time domain; (b) Absolute value of the transmitted and the received signal in the Fourier domain; (c) Channel impulse response; (d) Power delay profile. }
    \label{fig:channel_impulse}
\end{figure*} 

\subsection{Time and frequency analysis}
We compare the delay and channel impulse response from three neuron geometries. The simulation of the light path was for one dimension of 18 neurons that are linearly positioned along the wave propagation direction. In all cases, the receiver is located at a distance of 450 $\mu$m from the transmitter.
 
Fig.~\ref{fig:channel_impulse} shows both the time and frequency domain analysis. In our simulation, one femtosecond Gaussian light pulse is transmitted from a source which is located at 5 $\mu$m from the first cell on the path. The signal is analyzed at the receiver. The peak frequency of the transmitted signal is approximately 500 THz. From the ray tracing analysis, the delay can be characterized by integrating all the incident rays at 450 $\mu$m distance from the transmitter. 

Fig.~\ref{fig:channel_impulse} shows the channel impulse of the one-dimensional neuron cell and is obtained by FFT and IFFT as explained in Sec.~\ref{sec:time_frequency}. As shown in Fig.~\ref{fig:channel_impulse}, the channel impulse response exhibits correlation with the power delay profile in term of its peak magnitude of the received signal in time domain. The light propagating through the fusiform cells experiences higher delay compared to the other two cell types since the the light propagates mostly though the one dimensional array of neural tissue. In general, the speed of light in the interstitial tissue is faster than in the neuron because of the smaller refractive index. On the other hand, the light is absorbed less by the neuron due to its lower absorption coefficient, resulting in lower magnitude attenuation of the signal. The segmented time delays found in the pyramidal cells is caused by the ray leakages that occurs intermittently along the tissue. This phenomenon does not occur in two other type of cells since most of the ray propagation in two other cells are maintained along the straight path, even though there will be divergences and leakages. 

\begin{figure*}[t!]
    \centering
    \subfigure[Fusiform.]
        {
        \includegraphics[width=3in]{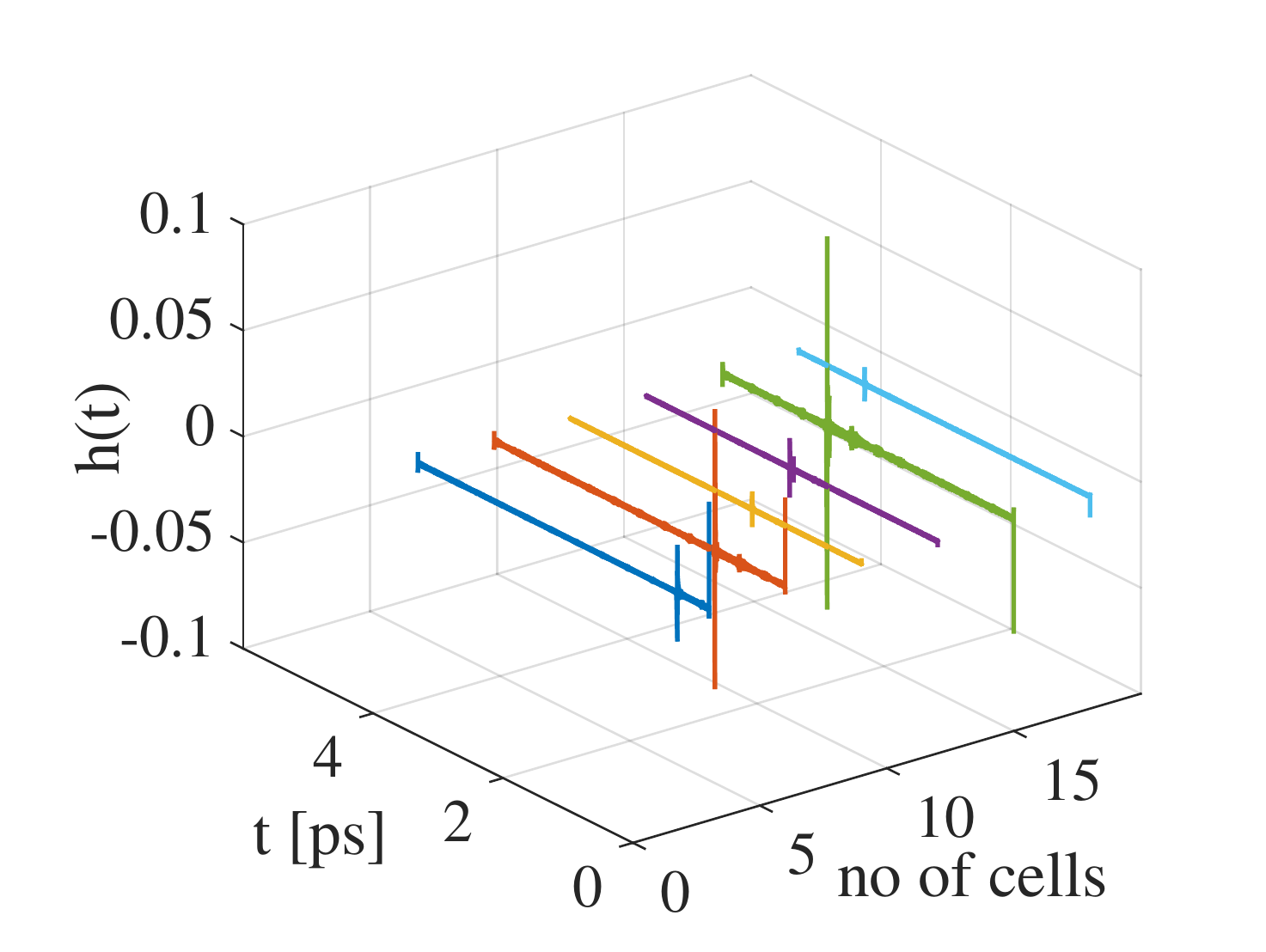}
        \label{fig:fusiform_cir}
        }
    ~
    \subfigure[Spherical.]
        {
        \includegraphics[width=3in]{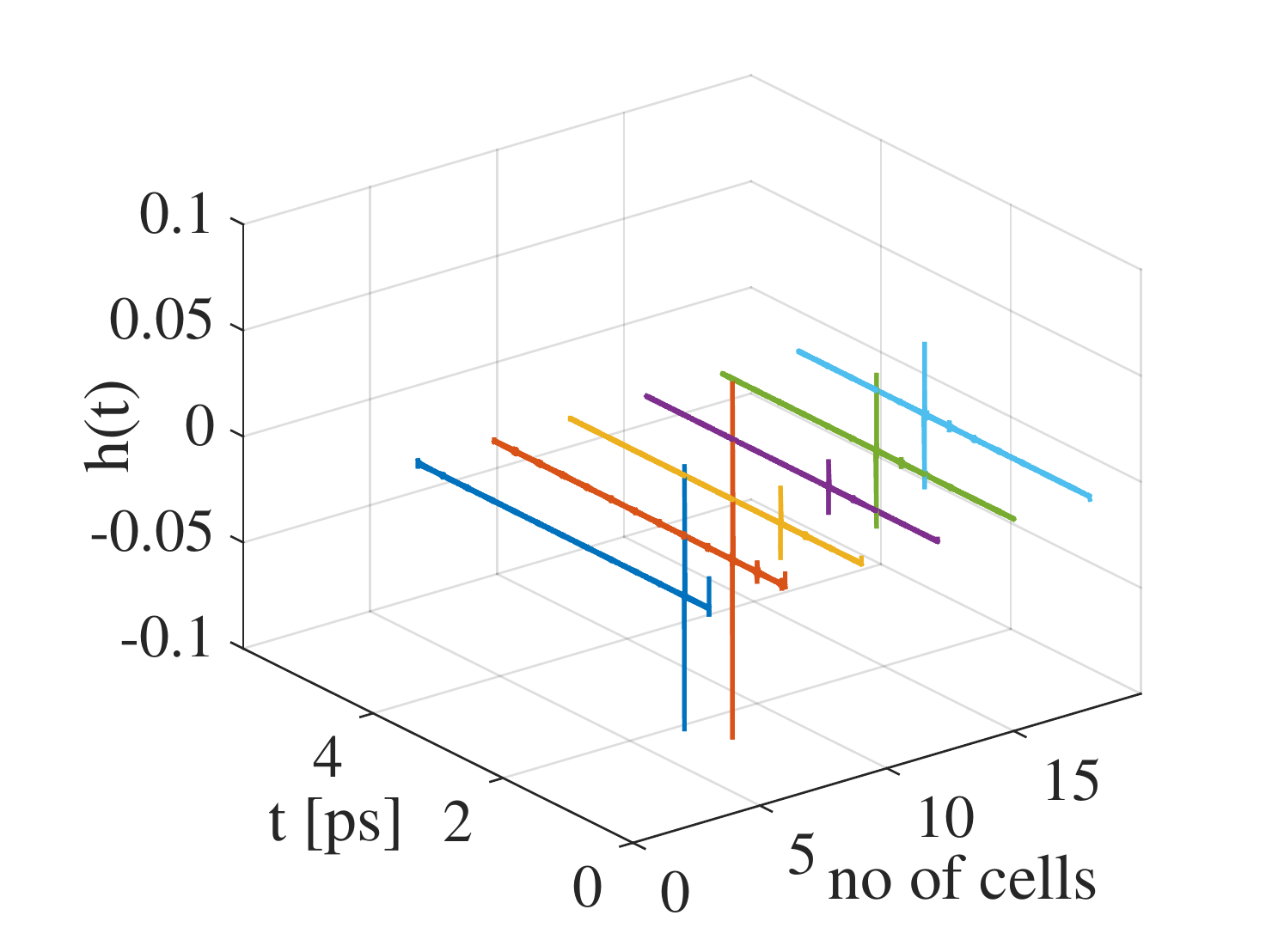}
        \label{fig:spherical_cir}
        }
    ~
    \subfigure[Pyramidal.]
        {
        \includegraphics[width=3in]{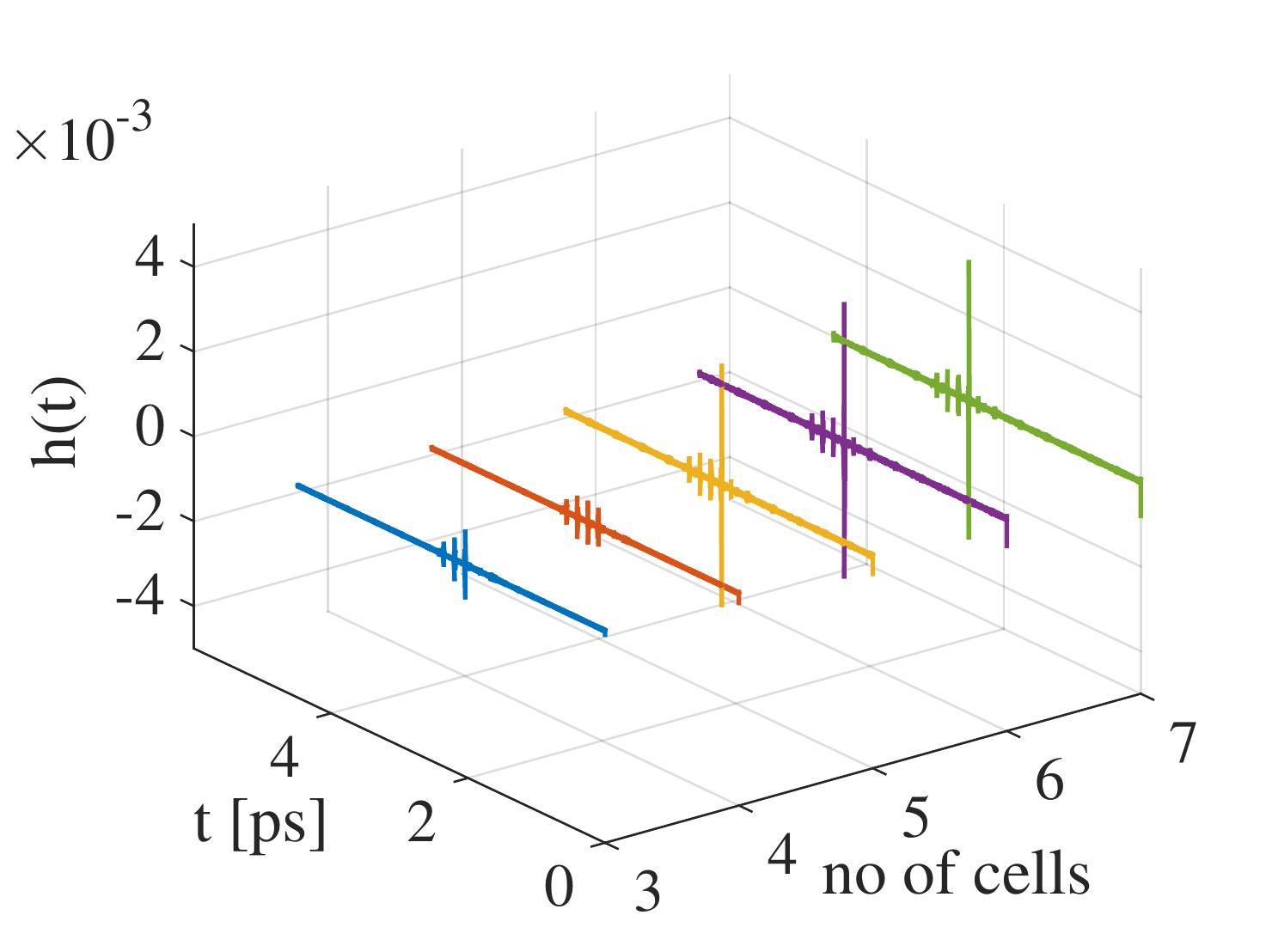}
        \label{fig:pyramidal_cir}
        }
    \caption{Channel impulse response $h(t)$ for three shapes of cells with variable number of cells along the propagation path.}
    \label{fig:channel_impulse_response}
\end{figure*}

Fig.~\ref{fig:channel_impulse_response} depicts the channel impulse response for the three cell shapes with respect to the number of cells along the propagation path. It is obvious that the number of cells has a  significant effect on the time of arrival of the signal, since the refractive index of the cell is higher resulting in slower light propagation velocity. The magnitude of the signal is dictated by the attenuation medium properties for all cell types and focusing parameter for both fusiform and spherical cells, which is determined in ~\eqref{eq:E_r}. The difference between the three cells is solely due to geometry which affects the distance of the focus point $foc(\theta_F, x_F)$, focusing parameter $\gamma$, and the ratio of total light propagation distance in the cell $\overline{d_a}$ and the interstitial tissue $\overline{d_e}$. The impulse response for the pyramidal cells increases gradually as the number of cell goes higher due to less traveled distance before the light is diverged leading to no focusing effect.

\subsection{Power characteristics on the receiver}
\begin{figure}[t!]
    \centering
    \includegraphics[width=\textwidth]{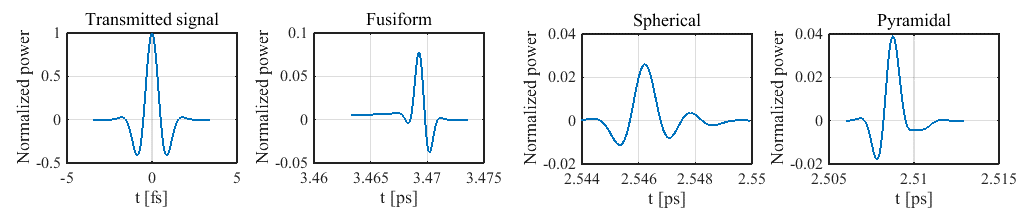} 
    \caption{The highest magnitude pulse shapes of the transmitted signal (gaussian short pulse) and the received signals after 450 $\mu$m propagation through one dimensional array of cells.}
    \label{fig:pulse_shape}
\end{figure}

On the receiver, the total received signal is the result of the superposition of the light rays that arrive at the receiver surface. Fig. \ref{fig:pulse_shape} shows the peak transmitted and received pulse shapes of different cell types as the result of superposition. Furthermore, the received signal power difference for those cells is mainly caused by the propagation distance ratio of brain tissue for each path and neuron and the geometrical gain for each cell shape.

\begin{figure}[b!]
    \centering
    \subfigure[Fusiform.]
        {
        \includegraphics[width=3in]{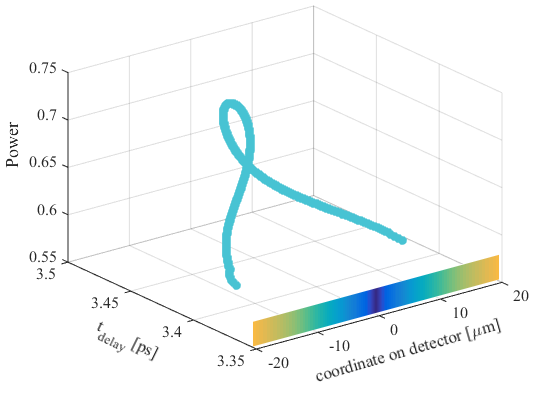}
        \label{fig:fusiform_detector}
        }
    ~
    \subfigure[Pyramidal.]
        {
        \includegraphics[width=3in]{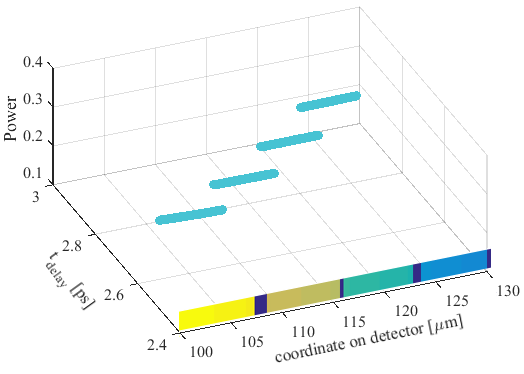}
        \label{fig:spherical_detector}
        }
    \caption{Normalized power of light ray when it arrives at a certain coordinate on a detector and the transmission delay.}
    \label{fig:detector_ray_coord}
\end{figure}

The previous results have shown that there are different propagation paths depending on the cell type, and this has an impact on how the detector on the receiver is designed. All the cells along the one-dimensional neural tissue contributes to the delay and attenuation of the light signal. Consequently, the geometry and the amount of the traversed cell also has an impact on the received power characteristics. Fig.~\ref{fig:detector_ray_coord} shows the received power characteristics on the detector of the receiver for both the fusiform and pyramidal cells. In the case of the fusiform cell's receiver detector, the width is 40 $\mu$m, while in the case of the pyramidal it is 30 $\mu$m. In the case of the fusiform cells (Fig.~\ref{fig:detector_ray_coord} (a)), the power concentration is located in the middle of the detector and this is in-line to the propagation direction where the cells are aligned. This is directly linked to the focusing phenomenon that occurs in both fusiform and spherical cells. In order to detect all the transmitted rays, the detector height should be equal to the height of the cell, $h_c$, which is 30 $\mu$m used in our simulation. On the other hand, different characteristics is observed for the pyramidal cells where gradual power change occurs in stages and it is also segmented (Fig.~\ref{fig:detector_ray_coord} (b)). While the height configuration of the detector used is 30 $\mu$m, this parameter is dependent on the cell configuration (the number of traversed cells). The segmented phenomenon occurs due to leaking rays effect discussed in the earlier section. Unlike the fusiform and spherical cells, which causes power concentration at the centre of the detector, in case of pyramidal cells, the light rays that traverses through more cells experiences less attenuation as well as deviations from the original propagation line. Therefore, for both fusiform as well as spherical, the detector should be positioned at the center of the propagation line. In the case of the pyramidal cells,  there are two ways for positioning the detector. As illustrated in Fig.~\ref{fig:ray_deviation}, there are both horizontal and vertical orientations. The horizontal orientation provides better benefit since the propagation distance can be minimized, but the detected intensity is higher than in the vertical orientation.

\section{Conclusion}
\label{sec:conclusion}
While light has been investigated for neural stimulation through the concept of optogenetics, this paper addresses the light propagation from a micron-scale light source that can be used for communication between the WiOptND devices. The analysis presented in this paper discussed the important factors the affect the light propagation through neurons and interstitial brain tissue are the medium optical properties ($\mu_a$ and $\mu_s$), as well as geometric structures of the cells. The cells investigated in this paper are fusiform, spherical, and pyramidal. An interesting effect is the distance and the number of the cells along the propagation axis, which affects the path loss as well as the geometrical gain.

The channel impulse response of light propagating along the neurons have an interesting behaviour. In the time domain, the delay of the simulated system can be observed when the light pulse is sufficiently short (femto second) since the delay is in pico second level. This means that for longer pulses, the delay is insignificant.

The time delays for the fusiform, spherical, and pyramidal are approximately 3.4, 2.5 and 2.7 ps, respectively. Additionally, radial-based geometries (fusiform and spherical) exhibit alternating high and low amplitude, while pyramidal tends to exhibit increasing amplitude signal level as the distance increases due to the path traversing through neurons more than the interstitial tissue. However, all shapes experience increasing delay as the distance increases. In terms of the Fourier domain, the propagated signal does not experience any change in its frequency range. Both the time and frequency analysis exhibit 35\%, 20\%, and 65\% attenuation in the signal power for the fusiform, spherical, and pyramidal cells, respectively. Moreover, the shape of the received signal is governed mainly by the geometrical shape of the cell where the diffraction causes the change in light directions for the pyramidal cell. The radial-based-geometry cell exhibited radial pattern in the power gradient at the receiver. In the case of the pyramidal cells, the position of the receiver is very important to obtain maximum light intensity for accurate detection. Our analysis found that the intensity of light intensity at the detector greatly varies across the area of the detector.

Analysis in this paper has shown that light propagation as a mode for communication between WiOptND implantable devices in the brain is a viable solution. The impulse response shows how the light propagation behaviour varies with the number of cells, and how this can impact on the area design of the detector. 


%

\appendices
\section{Ray tracing algorithm for fusiform cell}
This algorithm processes the optical properties of the medium ($n_c, n_t$), physical properties of the cell ($r_c, d_c$), and coordinates/direction of the incoming ray ($x_2, h_2, \theta_i$). It generates the focus angle/coordinates ($foc(\theta_F, x_F)$) and propagation direction/coordinates inside the cell ($l_i(x_3', h_3)$, $l_o(x_4, h_4)$, $\theta_i^{(2)}$). For a series of cells, the iterative execution of this algorithm is required.

\begin{algorithm}[ht]
  \caption{Ray tracing for fusiform cell}\label{alg:ray_function}
  \begin{algorithmic}[1]
    \Require
      \Statex $n_c, n_t$ (refractive indices of cell and tissue
      \Statex $r_c$ (the radius of the cell)
      \Statex $x_2$ (the x ray coordinate of the previous cell)
      \Statex $h_2$ (the radius of incoming illumination)
      \Statex $d_c$ (the distance between cells)
      \Statex $\theta_i$ (the angle of the incoming ray)
    \Ensure
      \Statex $foc(\theta_F, x_F)$ (the distance and angle of focus point
      \Statex $l_i(x_3', h_3)$ (the coordinate of the incoming ray)
 
      \Statex $l_o(x_4, h_4)$ (the coordinate of the outgoing ray)      
      \Statex $\theta_i^{(2)}$ (the ray propagation angle in the cell)
      \Statex
    
    \State \Call{Calculate} {$x_2'$} \Comment{$x_2'$ = $x_2$ measured from the 1$^{st}$ surface},
\[
x_2'= (d_c + 2r_c) - x_2
\]

    \algstore{part1}
  \end{algorithmic}
\end{algorithm}
\renewcommand{\thealgorithm}{1}
\begin{algorithm}
  \begin{algorithmic}[1]
    \algrestore{part1}   

	\State \Call{Calculate} {$x_3, h_3$} \Comment{$x_3, h_3$ = the coordinate where the ray hits the 1$^{st}$ surface},
\[
m_2= tan(180^o - \theta_i)
\]
\[
h_3 =
\begin{bmatrix}
    m_2^2+1 \\ 
    2m_2(h_2+ m_2 x_2')\\
    (h_2 + m_2 x_2')^2-r_c^2\\
\end{bmatrix}^T
\begin{bmatrix}
    x_3^2\\
    x_3\\
    1
\end{bmatrix}
\]
	\State \Call{Calculate}{$\theta_i^{(1)}$} \Comment{the incoming angle with respect to normal line of the 1$^{nd}$ surface}
\[
\theta_i'^{(1)} = arctan\Big(\frac{h_3}{|x_3|}\Big)-\theta_i^{(1)}
\]
	\State \Call{Calculate}{$\theta_o^{(1)}$} \Comment{the refracted angle due to 1$^{st}$ surface}
\[
\theta_o^{(1)} = arcsin\Big(\frac{n_t sin(\theta_i'^{(1)})}{n_c}\Big)
\]
	\State \Call{Calculate}{$x_3$, $\theta_o^{(1)}$} \Comment{with respect to 2$^{nd}$ surface}
\[
x_3' = 2r_c-(d_c+|x_3|)
\]
\[
\theta_i^{(2)} = \theta_i^{(1)} + (\theta_i'^(1) + \theta_o^{(1)})
\]
	\State \Call{Calculate}{$x_4, h_4$} \Comment{the coordinate where the ray hits the 2$^{nd}$ surface}
\[
m_3= tan(- \theta_i^{(2)})
\]
\[
h_4 =
\begin{bmatrix}
    m_3^2+1 \\ 
    2m_3(h_3+ m_3 x_3')\\
    (h_3 + m_3 x_3')^2-r_c^2\\
\end{bmatrix}^T
\begin{bmatrix}
    x_4^2\\
    x_4\\
    1
\end{bmatrix}
\]

	\State \Call{Calculate}{$\theta_o^{(2)}$} \Comment{the refracted angle due to 2$^{nd}$ surface}
\[
\theta_o^{(2)} = arcsin \Bigg( \frac{n_c}{n_t} sin\Big(arctan \Big(\frac{h_4}{x_4}\Big)+\theta_i^{(2)}\Big)\Bigg)
\]

\algstore{part2}
  \end{algorithmic}
\end{algorithm}
\begin{algorithm}
  \begin{algorithmic}[1]
    \algrestore{part2}
	\State \Call{Calculate}{$x_F$} \Comment{the focus distance}
\[
\theta_F = \theta_o^{(2)} - arctan \Big(\frac{h_4}{x_4}\Big)
\]
\[
m_4 = tan(\theta_F)
\]
\[
x_F = \frac{m_4 x_4 - h_4}{m_4}
\]
  \end{algorithmic}
\end{algorithm}


\section*{Acknowledgment}
The work has been supported by Science Foundation Ireland (SFI) CONNECT (13/RC/2077) Research Centre, as well as the Academy of Finland under Grant 284531.

\ifCLASSOPTIONcaptionsoff
  \newpage
\fi



%
%



\bibliographystyle{IEEEtran}
\bibliography{References}

\begin{thebibliography}{10}
\providecommand{\url}[1]{#1}
\csname url@samestyle\endcsname
\providecommand{\newblock}{\relax}
\providecommand{\bibinfo}[2]{#2}
\providecommand{\BIBentrySTDinterwordspacing}{\spaceskip=0pt\relax}
\providecommand{\BIBentryALTinterwordstretchfactor}{4}
\providecommand{\BIBentryALTinterwordspacing}{\spaceskip=\fontdimen2\font plus
\BIBentryALTinterwordstretchfactor\fontdimen3\font minus
  \fontdimen4\font\relax}
\providecommand{\BIBforeignlanguage}[2]{{%
\expandafter\ifx\csname l@#1\endcsname\relax
\typeout{** WARNING: IEEEtran.bst: No hyphenation pattern has been}%
\typeout{** loaded for the language `#1'. Using the pattern for}%
\typeout{** the default language instead.}%
\else
\language=\csname l@#1\endcsname
\fi
#2}}
\providecommand{\BIBdecl}{\relax}
\BIBdecl

\bibitem{akyildiz2010electromagnetic}
I.~F. Akyildiz and J.~M. Jornet, ``Electromagnetic wireless nanosensor
  networks,'' \emph{Nano Communication Networks}, vol.~1, no.~1, pp. 3--19,
  2010.

\bibitem{akyildiz2008nanonetworks}
I.~F. Akyildiz, F.~Brunetti, and C.~Bl{\'a}zquez, ``Nanonetworks: A new
  communication paradigm,'' \emph{Computer Networks}, vol.~52, no.~12, pp.
  2260--2279, 2008.

\bibitem{abbasi2018controlled}
N.~A. Abbasi, D.~Lafci, and O.~B. Akan, ``Controlled information transfer
  through an in vivo nervous system,'' \emph{Scientific reports}, vol.~8,
  no.~1, p. 2298, 2018.

\bibitem{ramezani2018impacts}
H.~Ramezani and O.~B. Akan, ``Impacts of spike shape variations on synaptic
  communication,'' \emph{IEEE transactions on nanobioscience}, vol.~17, no.~3,
  pp. 260--271, 2018.

\bibitem{veletic2016upper}
M.~Veleti{\'c}, P.~A. Floor, Y.~Chahibi, and I.~Balasingham, ``On the upper
  bound of the information capacity in neuronal synapses,'' \emph{IEEE
  Transactions on Communications}, vol.~64, no.~12, pp. 5025--5036, 2016.

\bibitem{johari2017nanoscale}
P.~Johari and J.~M. Jornet, ``Nanoscale optical wireless channel model for
  intra-body communications: Geometrical, time, and frequency domain
  analyses,'' \emph{IEEE Transactions on Communications}, vol.~66, no.~4, pp.
  1579--1593, 2017.

\bibitem{elayan2017photothermal}
H.~Elayan, P.~Johari, R.~M. Shubair, and J.~M. Jornet, ``Photothermal modeling
  and analysis of intrabody terahertz nanoscale communication,'' \emph{IEEE
  transactions on nanobioscience}, vol.~16, no.~8, pp. 755--763, 2017.

\bibitem{santagati2017software}
G.~E. Santagati and T.~Melodia, ``A software-defined ultrasonic networking
  framework for wearable devices,'' \emph{IEEE/ACM Transactions on Networking
  (TON)}, vol.~25, no.~2, pp. 960--973, 2017.

\bibitem{wirdatmadja2017wireless}
S.~A. Wirdatmadja, M.~T. Barros, Y.~Koucheryavy, J.~M. Jornet, and
  S.~Balasubramaniam, ``Wireless optogenetic nanonetworks for brain
  stimulation: Device model and charging protocols,'' \emph{IEEE transactions
  on nanobioscience}, vol.~16, no.~8, pp. 859--872, 2017.

\bibitem{noel2018distortion}
A.~Noel, D.~Makrakis, and A.~W. Eckford, ``Distortion distribution of neural
  spike train sequence matching with optogenetics,'' \emph{IEEE Transactions on
  Biomedical Engineering}, vol.~65, no.~12, pp. 2814--2826, 2018.

\bibitem{tuchin2015tissue}
V.~V. Tuchin, ``Tissue optics and photonics: biological tissue structures,''
  \emph{Journal of Biomedical Photonics \& Engineering}, vol.~1, no.~1, 2015.

\bibitem{fang2007intrinsic}
C.~Fang-Yen and M.~S. Feld, ``Intrinsic optical signals in neural tissues:
  Measurements, mechanisms, and applications,'' \emph{Acs Sym. Ser.}, vol. 963,
  pp. 219--235, 2007.

\bibitem{wirdatmadja2019analysis}
S.~Wirdatmadja, P.~Johari, A.~Desai, Y.~Bae, E.~K. Stachowiak, M.~K.
  Stachowiak, J.~M. Jornet, and S.~Balasubramaniam, ``Analysis of light
  propagation on physiological properties of neurons for nanoscale
  optogenetics,'' \emph{IEEE Transactions on Neural Systems and Rehabilitation
  Engineering}, vol.~27, no.~2, pp. 108--117, 2019.

\bibitem{rodriguez2013simulation}
S.~P. Rodr{\'\i}guez, R.~P. Jim{\'e}nez, B.~R. Mendoza, F.~J.~L. Hern{\'a}ndez,
  and A.~J.~A. Alfonso, ``Simulation of impulse response for indoor visible
  light communications using 3d cad models,'' \emph{EURASIP Journal on Wireless
  Communications and Networking}, vol. 2013, no.~1, p.~7, 2013.

\bibitem{scholkmann2013general}
F.~Scholkmann and M.~Wolf, ``General equation for the differential pathlength
  factor of the frontal human head depending on wavelength and age,''
  \emph{Journal of biomedical optics}, vol.~18, no.~10, p. 105004, 2013.

\bibitem{rulliere2005femtosecond}
C.~Rulliere \emph{et~al.}, \emph{Femtosecond laser pulses}.\hskip 1em plus
  0.5em minus 0.4em\relax Springer, 2005.

\bibitem{wirdatmadja2018light}
S.~Wirdatmadja, P.~Johari, S.~Balasubramaniam, Y.~Bae, M.~K. Stachowiak, and
  J.~M. Jornet, ``Light propagation analysis in nervous tissue for wireless
  optogenetic nanonetworks,'' in \emph{Optogenetics and Optical Manipulation
  2018}, vol. 10482.\hskip 1em plus 0.5em minus 0.4em\relax International
  Society for Optics and Photonics, 2018, p. 104820R.

\bibitem{levinson1926refractometric}
A.~Levinson and A.~Serby, ``The refractometric and viscosimetric indexes of
  cerebrospinal fluid,'' \emph{Archives of Internal Medicine}, vol.~37, no.~1,
  pp. 144--150, 1926.

\bibitem{yaroslavsky2002optical}
A.~Yaroslavsky, P.~Schulze, I.~Yaroslavsky, R.~Schober, F.~Ulrich, and
  H.~Schwarzmaier, ``Optical properties of selected native and coagulated human
  brain tissues in vitro in the visible and near infrared spectral range,''
  \emph{Physics in Medicine \& Biology}, vol.~47, no.~12, p. 2059, 2002.

\bibitem{bosschaart2014literature}
N.~Bosschaart, G.~J. Edelman, M.~C. Aalders, T.~G. van Leeuwen, and D.~J.
  Faber, ``A literature review and novel theoretical approach on the optical
  properties of whole blood,'' \emph{Lasers in medical science}, vol.~29,
  no.~2, pp. 453--479, 2014.

\end{thebibliography}

%




\end{document}